\documentclass[aps,prl,twocolumn,superscriptaddress,showpacs,floatfix,nofootinbib]{revtex4}
\usepackage{amsmath,graphicx}
\newcommand{\mean}[1]{\left\langle #1 \right\rangle}
\begin{document}

\title{Collective flow in event-by-event partonic transport plus
  hydrodynamics hybrid approach}

\author{Rajeev S. Bhalerao} 
\affiliation{Department of Theoretical Physics, Tata Institute of
  Fundamental Research, Homi Bhabha Road, Mumbai 400005, India}
\author{Amaresh Jaiswal} 
\affiliation{GSI, Helmholtzzentrum f\"ur Schwerionenforschung,
Planckstrasse 1, D-64291 Darmstadt, Germany}
\author{Subrata Pal} 
\affiliation{Department of Nuclear and Atomic Physics, Tata Institute
  of Fundamental Research, Homi Bhabha Road, Mumbai 400005, India}

\date{\today}

\begin{abstract}
Complete evolution of the strongly interacting matter formed in
ultrarelativistic heavy-ion collisions is studied within a coupled
Boltzmann and relativistic viscous hydrodynamics approach. For the
initial nonequilibrium evolution phase, we employ a multiphase
transport (AMPT) model that explicitly includes event-by-event
fluctuations in the number and positions of the participating nucleons
as well as of the produced partons with subsequent parton
transport. The ensuing near-equilibrium evolution of quark-gluon and
hadronic matter is modeled within the (2+1)-dimensional relativistic
viscous hydrodynamics.
We probe the role of parton dynamics in generating and maintaining the
spatial anisotropy in the preequilibrium phase. Substantial spatial
eccentricities $\varepsilon_n$ are found to be generated in the
event-by-event fluctuations in parton production from initial
nucleon-nucleon collisions. For ultracentral heavy-ion collisions,
the model is able to explain qualitatively the unexpected hierarchy of
the harmonic flow coefficients
$v_n(p_T)~(n=2-6)$ observed at energies currently available at the CERN 
Large Hadron Collider (LHC).
We find that the results for 
$v_n(p_T)$ are rather insensitive to the variation (within a range)
of the time of switchover from AMPT parton transport to hydrodynamic
evolution.
The usual Grad and the recently proposed Chapman-Enskog-like
(nonequilibrium) single-particle distribution functions are found to
give very similar results for $v_n~(n=2-4)$.
The model describes well both the BNL Relativistic Heavy Ion Collider
 and LHC data for $v_n(p_T)$ at
various centralities, with a constant shear viscosity to entropy
density ratio of 0.08 and 0.12, respectively.
The event-by-event distributions of $v_{2,3}$ are in good agreement
with the LHC data for midcentral collisions. The linear response
relation $v_n = k_n \varepsilon_n$ is found to be true for $n=2,3$,
except at large values of $\varepsilon_n$, where a larger value of
$k_n$ is required, suggesting a small admixture of positive nonlinear
response even for $n=2,3$.

\end{abstract}



\maketitle

\section{Introduction}

High-energy heavy-ion collision studies at the BNL Relativistic Heavy Ion
Collider (RHIC) \cite{Adams:2005dq,Adcox:2004mh} and the CERN Large Hadron
Collider (LHC) \cite{ALICE:2011ab,ATLAS:2012at,Chatrchyan:2013kba}
have firmly established the formation of a strongly interacting
quark-gluon plasma (QGP) close to local thermodynamic equilibrium.
Evidence for this is provided by the hydrodynamical
analyses of collective flow, which require an extremely small shear
viscosity to entropy density ratio $\eta/s$
\cite{Romatschke:2007mq,Song:2007ux}. The
precise extraction of $\eta/s$ depends crucially on the adequate
knowledge of the initial-state dynamics, the lack of which, at present,
represents the largest uncertainty in the hydrodynamic modeling of
heavy-ion collisions
\cite{Steinheimer:2007iy,Holopainen:2010gz,Schenke:2010rr,
  Schenke:2011bn,Qiu:2011iv}.

Smooth (nonfluctuating) initial energy-density distributions based on
the Glauber or color-glass-condensate (CGC) models have been used
successfully to describe the elliptic flow $v_2$, albeit with
differing values of $\eta/s$ \cite{Hirano:2005xf}. On the other hand,
event-by-event fluctuations in the initial configuration of the system
are solely responsible for the (rapidity-even) odd flow harmonics,
$v_3,~v_5$ \cite{Alver:2010gr}. The fluctuations are also important in
explaining the double-peaked structure in the final dihadron
correlations across large rapidities
\cite{Abelev:2008ac,Adare:2008ae, ALICE:2011ab,ATLAS:2012at}.

The Monte Carlo method has been invoked to simulate event-by-event
geometric fluctuations in the number and positions of the participant
nucleons (MC-Glauber)
\cite{Holopainen:2010gz,Qiu:2011iv,Schenke:2010rr,
  Schenke:2011bn} or of intrinsic gluons (MC-CGC)
\cite{Qiu:2011iv}.  Both methods give rise to ``lumpy"
initial conditions and are able to reproduce elliptic flow data at
RHIC and LHC \cite{Romatschke:2007mq, Qiu:2011iv}. MC-CGC yields a
relatively larger initial eccentricity and and hence requires larger
viscous damping ($\eta/s$) to explain the elliptic flow data .
However, it underpredicts the triangular flow data \cite{Qiu:2011hf}.
Most of these model studies ``switch on'' hydrodynamics at a specified
early time, ignoring the preequilibrium dynamical evolution towards
the hydrodynamic regime. Indeed, various ansatze and parametrizations
have been adopted to describe the initial distribution of the unknown
kinetic and thermodynamic quantities in these models.

In a more recent study, the preequilibrium dynamics was approximately
accounted for by combining an impact-parameter-dependent saturation model
with the classical Yang-Mills description of the flowing glasma fields
\cite{Gale:2012rq}. The model considered the color charge fluctuations
inside the nucleon as well and could describe remarkably well the
transverse-momentum ($p_T$)-dependent and $p_T$-integrated flow
harmonics $v_n~ (n=1-5)$ data at LHC
\cite{ALICE:2011ab,ATLAS:2012at}. A fully dynamical simulation which
included solutions to anti-de Sitter/conformal field theory (AdS/CFT)
for the preequilibrium stage and the
standard viscous hydrodynamics for the equilibrium stage was recently
developed \cite{vanderSchee:2013pia}. Preequilibrium dynamics has
also been accounted for in the ultrarelativistic quantum molecular dynamics 
(UrQMD) string dynamics model
\cite{Steinheimer:2007iy}. These studies underscore the importance of
including the inherent fluctuations at both nucleonic and partonic
levels as well as the initial-state preequilibrium dynamics.

The aim of the present paper is to study various aspects of the
hydrodynamic flow in a model with an initial state that includes
fluctuations at the nucleonic as well as subnucleonic levels and
which moreover considers preequilibrium dynamics of partonic
matter. For this purpose we have used a multiphase transport (AMPT)
model \cite{Lin:2004en} that consistently incorporates these
fluctuations in the preequilibrium phase. For the subsequent
near-equilibrium evolution, the standard (2+1)-dimensional [(2+1)D] 
viscous hydrodynamics
code VISH2+1 \cite{Song:2007ux} is
used until the matter reaches freeze-out.

\section{The model}

The version of AMPT used in this paper has the initial conditions
based on the HIJING 2.0 model \cite{Deng:2010mv,Pal:2012gf}, which
employs the Glauber model with Woods-Saxon nuclear distribution to
determine the nucleon configuration in an event. While the soft (binary)
nucleon-nucleon collisions lead to string excitations, the hard
collisions produce uncorrelated minijet partons. Fluctuations in the
minijet parton multiplicity follow the Poisson distribution where the
average multiplicity (determined by the minijet cross section)
increases strongly with the center-of-mass energy
\cite{Wang:1991hta}. We have employed the string-melting version of
AMPT \cite{Lin:2004en} for the initial partonic state, which
consistently incorporates these fluctuations and describes better the
final collective behavior of hadrons. In this version the strings melt
into their constituent quarks and antiquarks, whose positions are
determined randomly within the string, which is yet another source of
fluctuations. The scatterings among these quarks and minijet partons
are treated with Zhang's parton cascade (ZPC) with a parton-parton elastic cross
section of 1.5 mb.

\begin{figure}[t]
 \begin{center}
  \scalebox{.5}{\includegraphics{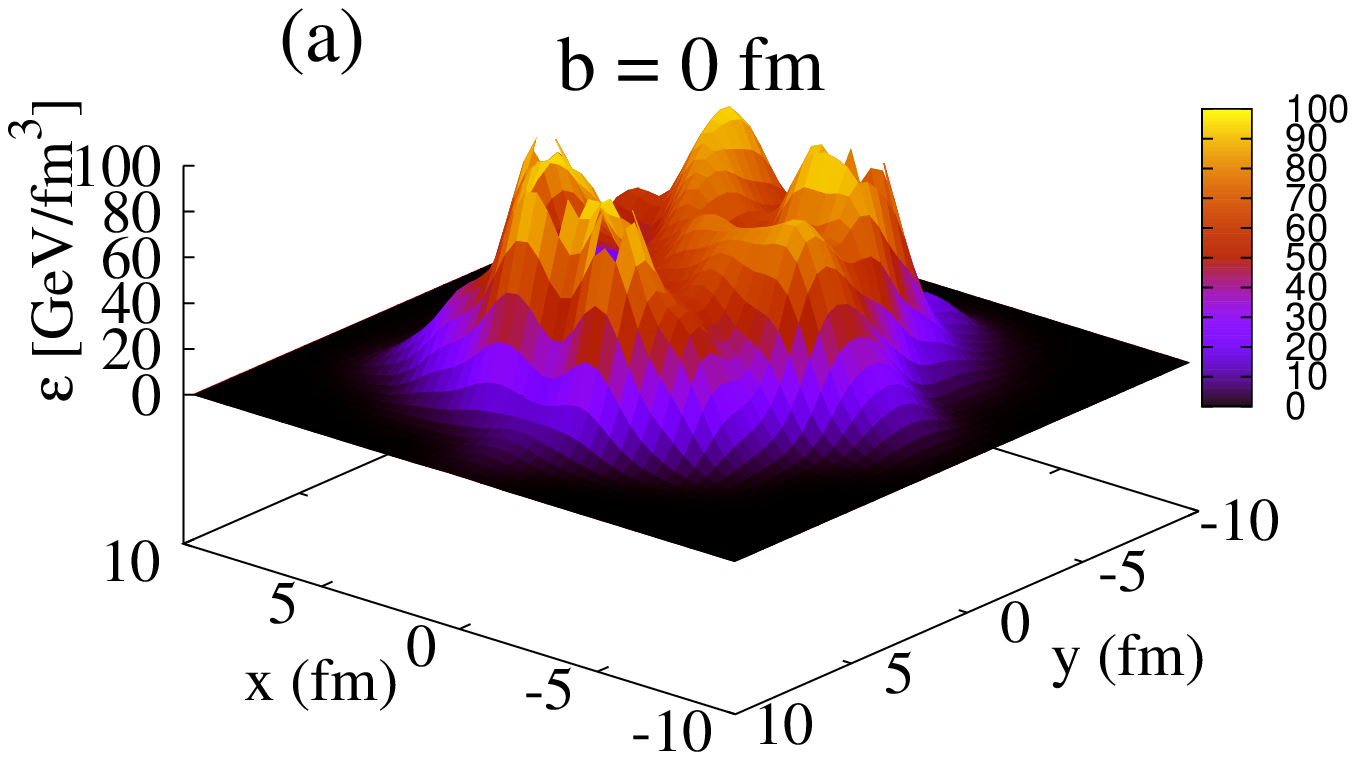}}\hfil
  \scalebox{.5}{\includegraphics{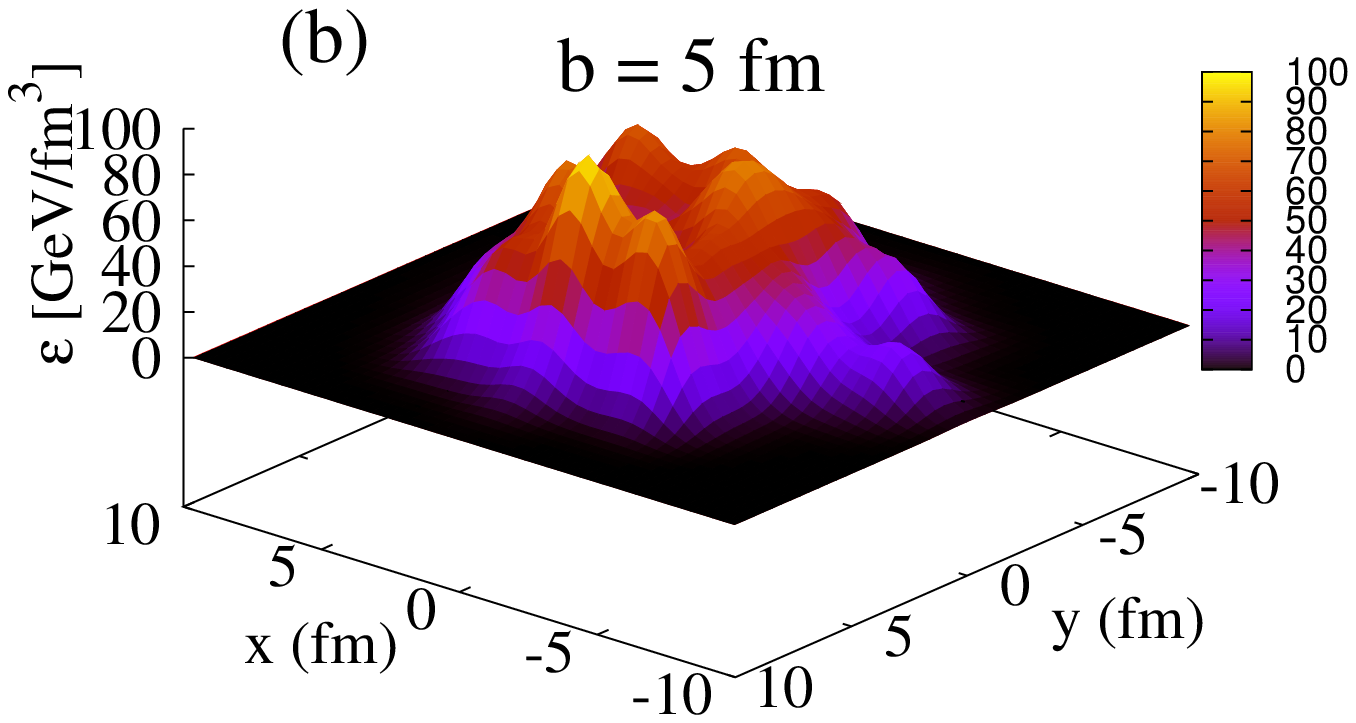}}
 \end{center}
 \vspace{-0.6cm}
 \caption{(Color online) Energy-density distribution in the AMPT model at the
   switchover time $\tau_{\rm sw}=0.4$ fm/$c$ in the transverse $xy$
   plane, with the smearing parameter $\sigma=0.8$ fm. The results are
   for (a) central ($b=0$ fm) and (b) noncentral ($b=5$ fm) Pb+Pb collisions
   at a center-of-mass energy $\sqrt{s_{NN}}=2.76$ TeV.}
 \label{exy}
\end{figure}

At the switchover time $\tau_{\rm sw}$, from the microscopic AMPT
model to the macroscopic viscous VISH2+1 hydrodynamic model
\cite{Song:2007ux}, the local energy-momentum density,
baryon number density, and the flow velocity in an event can be 
estimated from the positions and momenta 
of the formed partons at midrapidity: We use the space-time
rapidity window $[-1,1]$. To remove the possible
numerical instabilities in the ensuing hydrodynamic evolution, each
pointlike parton in AMPT is smeared with a 2D Gaussian
distribution in the transverse plane
\cite{Steinheimer:2007iy,Holopainen:2010gz}. Energy
density in the local rest frame then becomes
\begin{equation}\label{smear}
\epsilon(x,y) = \frac{N}{2\pi\sigma^2 \tau_{\rm sw}} \sum_i E'_i
\exp \left[ - \frac{(x-x_i)^2 + (y-y_i)^2}{2\sigma^2} \right] ,
\end{equation}
where ($x_i,\,y_i$) are the transverse coordinates and $E'_i$ is the
energy of the $i$th parton in the local rest frame.
The parameter $N$ is introduced to account for the various
uncertainties inherent in the present approach, such as those owing to
our prescription to match the (3+1)D microscopic description with the
(2+1)D macroscopic one, at the selected switchover time $\tau_{\rm sw}$ and
the smearing parameter $\sigma$, any other missing preequilibrium
physics, lack of hadronic after-burner, etc. The value of $N$ was
adjusted to describe the most central $p_T$ spectra. $N$ was found to
be of the order of unity: 1.38 at RHIC and 1.24 at LHC. The same value was
used at all other centralities. A similar normalization constant has
been used in Refs.
\cite{Steinheimer:2007iy,Holopainen:2010gz,Gale:2012rq,Pang:2012he}.
The Gaussian width
is set at $\sigma =0.8$ fm. Larger (smaller) values of $\sigma$ lead
to smaller (larger) fluctuations and hence reduced (enhanced)
magnitudes of the odd flow harmonics, $v_n$. 
Final numerical results were found to be 
insensitive to the choice of the initial transverse flow velocity
\cite{Gale:2012rq}; hence, we have set it equal to zero.
A detailed study of the effects of the initial flow velocity was performed in
Ref. \cite{Pang:2012he}.

Figure \ref{exy} shows the energy-density profiles in the transverse
plane, calculated in the AMPT model, for Pb+Pb collisions at
$\sqrt{s_{NN}}=2.76$ TeV, in a central (impact parameter $b=0$ fm) and a
noncentral ($b=5$ fm) event. The snapshots are taken at the
switchover time $\tau_{\rm sw}=0.4$ fm/$c$. As expected, the high-energy-density 
zone is of wider spatial extent in the central collision than
in the noncentral collision. We further notice that these
distributions are azimuthally anisotropic and also contain several
peaks which correspond to local maxima or ``hot spots"
\cite{Ma:2010dv}, which result in enhanced pressure gradients.

The hydrodynamic evolution is governed by the conservation equations
for particle current, $\partial_\mu N^\mu=0$, and the energy-momentum
tensor, $\partial_\mu T^{\mu\nu} =0$, where
\begin{align}\label{NTD}
N^\mu &= nu^\mu + n^\mu,  \nonumber\\
T^{\mu\nu} &= \epsilon u^\mu u^\nu - P\Delta^{\mu \nu} + \pi^{\mu\nu}.
\end{align}
Here $P, n, \epsilon$ are, respectively, hydrodynamic pressure, number
density, and energy density, and $\Delta^{\mu\nu}=g^{\mu\nu}-u^\mu u^\nu$
is the projection operator on the three-space orthogonal to the
hydrodynamic four-velocity $u^\mu$ defined in the Landau frame:
$T^{\mu\nu} u_\nu=\epsilon u^\mu$. The dissipative quantities, viz.,
the particle diffusion current and the shear pressure tensor, are
denoted by $n^\mu$ and $\pi^{\mu\nu}$, respectively. In this work, we
set the net baryon number current, the bulk viscous pressure, and the
initial shear viscous tensor to be zero.

We have employed the s95p-PCE equation of state (EoS)
\cite{Huovinen:2009yb}, which is obtained from fits to lattice data
for crossover transition and matches a realistic hadron resonance
gas model at low temperatures $T$, with partial chemical equilibrium
(PCE) of the hadrons for temperatures below $T_{\rm PCE} \approx 165$
MeV.

\begin{figure}[t]
 \begin{center}
  \scalebox{0.35}{\includegraphics{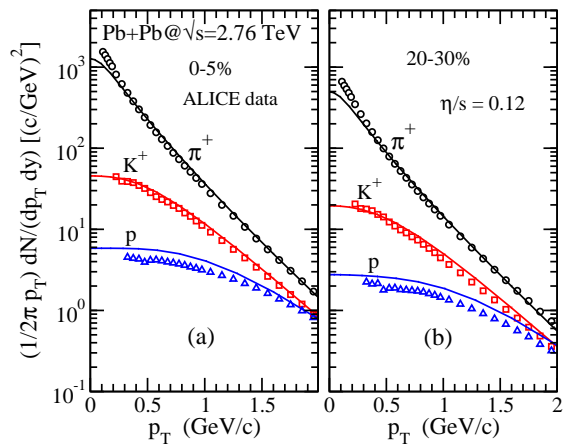}}
 \end{center}
 \vspace{-0.4cm}
 \caption{(Color online) Transverse-momentum spectra of $\pi^+$, $K^+$, and $p$ for
   two centrality ranges, $0-5$\% and $20-30$\%, in Pb+Pb collisions at
   $\sqrt{s_{NN}}=2.76$ TeV. The symbols represent ALICE data
   \cite{Abelev:2013vea} at midrapidity and the lines correspond to
   AMPT+Hydro calculations.}
 \label{pt_piKp_LHC}
\end{figure}

The hydrodynamic evolution is continued until each fluid cell reaches a
decoupling temperature of $T_{\rm dec} =120$ MeV. The hadronic spectra
are obtained at this temperature using the Cooper-Frye freeze-out
prescription \cite{Cooper:1974mv}
\begin{equation}\label{CF}
\frac{dN}{d^2p_TdY} = \frac{g}{(2\pi)^3} \int p_\mu d\Sigma^\mu f(x,p),
\end{equation}
where $Y$ is the rapidity, $g$ is the degeneracy factor, $p^\mu$ is
the particle four-momentum, $d\Sigma^\mu$ represents the element of
the 3D freeze-out hypersurface and $f(x,p)$ is the
nonequilibrium phase-space distribution function at freeze-out, which
can be written as a small deviation from the equilibrium distribution
function $f_0$, i.e., $f=f_0+\delta f$. Unless otherwise mentioned, we
use the viscous correction form corresponding to Grad's
14-moment approximation \cite{Grad},
\begin{equation}\label{Grad}
 \delta f = \frac{f_0 \tilde f_0}{2(\epsilon+P)T^2}\, p^\alpha p^\beta \pi_{\alpha\beta},
\end{equation}
where corrections up to second order in momenta are present, and
$\tilde f_0 \equiv 1-r f_0$, with $r=1,-1,0$ for Fermi, Bose, and
Boltzmann gases, respectively. Resonances of masses up to 2.25 GeV are
included in the calculations (so as to be consistent with the s95p-PCE
EoS). Results presented here include resonance decays; this tends to
reduce the anisotropic flow especially at low transverse momenta.

\section{Results and discussions}

Figure \ref{pt_piKp_LHC} shows the transverse-momentum spectra of
pions, kaons, and protons in the VISH2+1 calculation with AMPT
fluctuating initial conditions in the $0-5$\% and $20-30$\% central
Pb+Pb collisions at $\sqrt{s_{NN}}=2.76$ TeV in comparison with the
ALICE data at midrapidity \cite{Abelev:2013vea}. To connect the
centrality $c$ to the impact parameter $b$, we use the empirical
relation $c = \pi b^2 /\sigma$ \cite{Xu:2011fe}, with the
nucleus-nucleus total inelastic cross section $\sigma = 784$ fm$^2$
calculated from the Glauber model.
The results are obtained with
200 events, with a shear viscosity to entropy density ratio $\eta/s
=0.12$ in the VISH2+1 calculation. Here and in subsequent figures, 
unless stated otherwise, 
error bars for the relevant $p_T$ range are comparable to the symbol
sizes. We find that the spectra for
$\pi^+$ and $K^+$ from the full AMPT+Hydro simulations are in good
overall agreement with the experimental data. However, the proton
yield is overpredicted in the model possibly owing to the neglect of
the hadron spectrum above 2.25 GeV and final-state hadronic
rescatterings, particularly $p {\bar p}$ annihilations.  Incidentally,
this proton excess is in line with the ``proton anomaly'' seen in the
statistical hadronization model fits to the LHC data at this energy
\cite{Stachel:2013zma}.

\begin{figure}[t]
 \begin{center}
  \scalebox{0.32}{\includegraphics{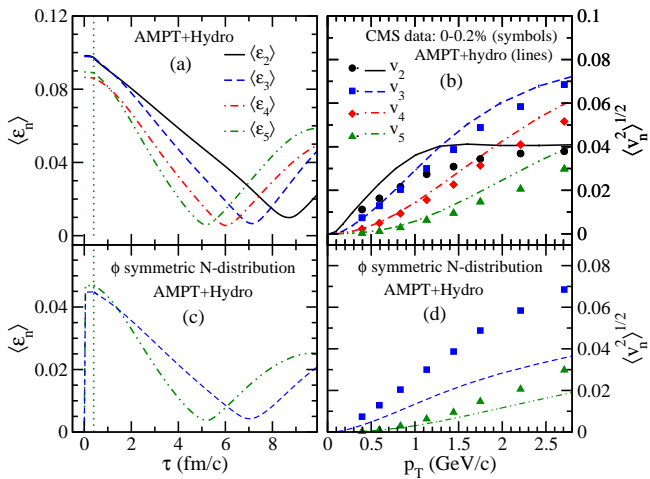}}
 \end{center}
 \vspace{-0.4cm}
 \caption{(Color online) Time evolution of event-averaged eccentricity $\langle
   \varepsilon_n\rangle$ and $p_T$ dependence of root-mean-square
   anisotropic flow coefficients $\langle v_n^2\rangle^{1/2}$ for
   charged hadrons in the AMPT+Hydro calculations for head-on ($b=0$)
   Pb+Pb collisions at $\sqrt{s_{NN}}=2.76$ TeV. The vertical line at
   $\tau=0.4$ fm/$c$ corresponds to the switchover time. Results are for
   the nucleon distributions in the colliding nuclei, which are
   azimuthally asymmetric (a),(b) and symmetric (c),(d); see text for
   details. The symbols represent the CMS data \cite{CMS:2013bza} for
   $v_n \{2\}$ obtained from two-particle correlations in the 0-0.2$\%$
   ultracentral Pb+Pb collisions at LHC.}
 \label{init_LHC}
\end{figure}

The $n$th harmonic participant eccentricity in the AMPT model can be
estimated from the transverse positions of the produced partons as
\cite{Petersen:2010cw}
\begin{equation}\label{eccn}
 \varepsilon_n = \frac{\sqrt{ \langle r^n \cos(n\phi) \rangle^2 + 
 \langle r^n \sin(n\phi) \rangle^2} } {\langle r^n\rangle}.
\end{equation}
Here $(r,\phi)$ are the polar coordinates of the partons 
(or of the nucleons at $\tau=0$) in the
transverse plane, and the average $\langle \cdots\rangle$ in a given
event is obtained by using the smeared local energy density as the
weight. For the calculation of $\varepsilon_n$ during the
hydrodynamical evolution, $(r,\phi)$ refer to the cell coordinates.
Figure \ref{init_LHC}(a) depicts the time evolution of event-averaged
eccentricities in central $(b=0)$ collisions. For these collisions,
eccentricities arise purely from event-to-event fluctuations of
particle positions. The higher-harmonic eccentricities are found to
drop faster with time for $\tau > \tau_{\rm sw} = 0.4$ fm/$c$. The
minima correspond to the change in the sign of $\varepsilon_n$ at late
times.

We recall that the fluctuations arise at both nucleonic and partonic
levels. To isolate the contribution of partonic fluctuations,
we consider the nucleon configuration obtained via Monte Carlo Glauber
in one quadrant of the overlap zone and replicate it in all other
quadrants, by imposing reflection symmetry with respect to the $x$ and
$y$ axes, in each event. This results in a nucleon distribution which
is approximately azimuthally symmetric in the initial state. As seen
in Fig. \ref{init_LHC}(c), this causes 
eccentricities $\mean{\varepsilon_n}$ to
vanish at $\tau=0$. However, the subsequent production of partons
through (semi-)hard nucleon-nucleon collisions results in a sizable
and rapid generation of $\langle \varepsilon_n \rangle$.
For even harmonics (not shown in the figure) the rise is higher than that
for odd harmonics, owing to the initial azimuthal
symmetry.

The flow harmonics $v_n$ are generated by conversion of the initial
spatial anisotropy into momentum anisotropy owing to unequal pressure
gradients in the hydrodynamic evolution. To calculate the $v_n$ in the
event-by-event fluctuating AMPT plus VISH2+1 hybrid model and compare
it with the LHC data, we first estimate for each harmonic the
event-plane angle $\Psi_n$ as in
\cite{Schenke:2010rr,Schenke:2011bn,Petersen:2010cw}
\begin{align}
\Psi_n = \frac{1}{n} \arctan 
\frac{\langle \sin (n\phi)\rangle} {\langle \cos (n\phi)\rangle}, 
\label{Psin}
\end{align}
where $\phi$ is the azimuthal angle of the outgoing hadron momentum.
The flow coefficients are then determined using
\begin{align}
v_n(p_T) &= \langle \cos [n(\phi-\Psi_n)] \rangle \nonumber\\
&\equiv \frac{\int d\phi \, \cos [n(\phi-\Psi_n)] \ dN/(dY \: p_T \: dp_T \: d\phi)}
{\int d\phi \, dN/(dY \: p_T \: dp_T \: d\phi)}, \label{vnpt}
\end{align}
where $dN/(dY \: p_T \: dp_T \: d\phi)$ is the particle spectrum. 
As in Eq. (\ref{eccn}), the angular brackets in Eqs. (\ref{Psin}) and (\ref{vnpt})
also refer to 
averaging over all the relevant particles in a given event. In
the present (2+1)D (longitudinal boost invariant) viscous
hydrodynamic simulation, the dependence on the rapidity $Y$ is
suppressed. We calculate the root-mean-square value of the flow
harmonics, $\langle v_n^2(p_T)\rangle^{1/2}$, over the entire event sample 
and compare it with the $v_n$ data at RHIC and LHC.
We recall that
for the $v_n\{2\}$ data obtained with the two-particle correlation method or
the $v_n\{EP\}$ data obtained with the event-plane method, the 
root-mean-square (rms) $v_n$ is
the most appropriate quantity to compare with \cite{Bhalerao:2006tp}.

Figures \ref{init_LHC}(b) and \ref{init_LHC}(d) show the $p_T$ dependence of the
rms $v_n$
for charged hadrons at small $p_T$. For the azimuthally symmetric
nucleon distribution, $\epsilon_3$ and $\epsilon_5$ 
originating purely from parton
fluctuations translate into sizable fractions ($\sim 50\%$) of the
total $v_3$ and $v_5$, respectively; see Fig. \ref{init_LHC}(d).

\begin{figure}[t]
 \begin{center}
  \scalebox{0.34}{\includegraphics{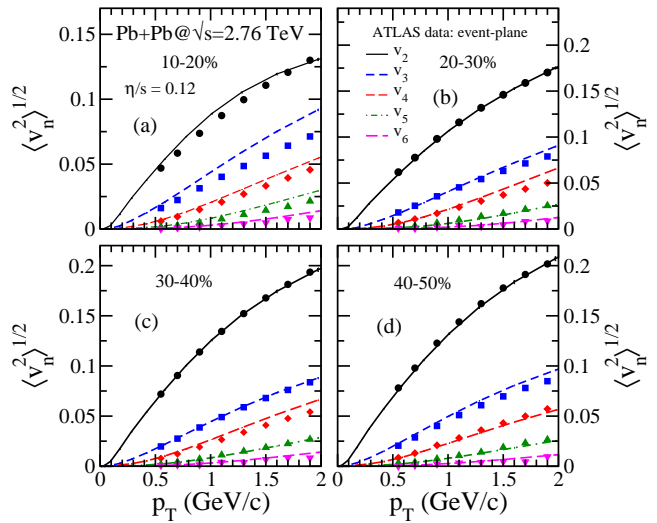}}
 \end{center}
 \vspace{-0.4cm}
 \caption{(Color online) Transverse-momentum dependence of the rms
   anisotropic flow coefficients $\langle v_n^2(p_T)\rangle^{1/2}$ of
   charged hadrons calculated at various centralities in Pb+Pb
   collisions at $\sqrt{s_{NN}}=2.76$ TeV in the AMPT+Hydro hybrid approach
   (lines) with $\eta/s=0.12$ as compared to the ATLAS data
   \cite{ATLAS:2012at} (symbols) based on the event-plane method.}
 \label{vnpt_b1234_LHC}
\end{figure}

Figure \ref{init_LHC}(b) also displays an intriguing feature observed
in the CMS data for $v_n(p_T)$ in ultracentral ($0-0.2\%$) Pb+Pb
collisions at $\sqrt{s_{NN}}=2.76$ TeV: Elliptic flow $v_2(p_T)$ has the
largest magnitude up to $p_T \simeq 1$ GeV/$c$ and thereafter it
flattens out, whereas the higher harmonics continue to rise and become
successively larger than $v_2$ with increasing $p_T$. This is in
contrast to what is observed at higher centralities; see
Fig. \ref{vnpt_b1234_LHC}, for instance. Our hybrid-model
calculations are in qualitative agreement with the CMS data over the
entire $p_T$ range. Better understanding of ultracentral collisions
is still a challenge for the theory \cite{Jia:2014jca,Denicol:2014ywa}.

\begin{figure}[t]
 \begin{center}
  \scalebox{0.38}{\includegraphics{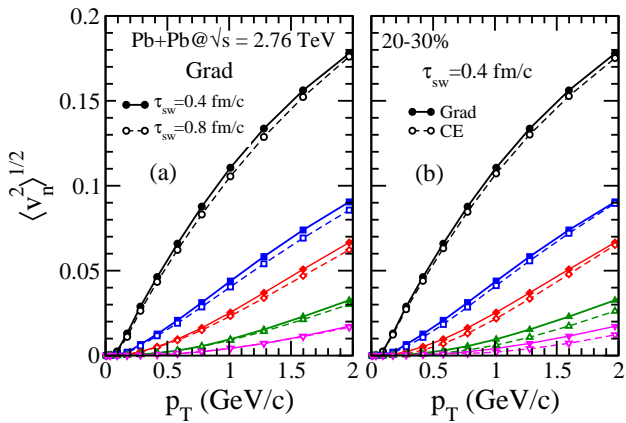}}
 \end{center}
 \vspace{-0.4cm}
 \caption{(Color online) Root-mean-square anisotropic flow coefficients, $\langle
   v_n^2(p_T)\rangle^{1/2}$ (for $n=$ 2-6, top to bottom), for
   charged hadrons in the AMPT+Hydro calculations in $20-30$\% central
   Pb+Pb collisions at $\sqrt{s_{NN}}=2.76$ TeV and $\eta/s=0.12$. The
   results are (a) for two different switchover times, $\tau_{\rm
     sw}=0.4$ fm/$c$ (solid lines) and $\tau_{\rm sw}=0.8$ fm/$c$ (dashed
   lines); and (b) for two different forms of $\delta f$ in the
   freeze-out prescription, viz., Grad's approximation
   [Eq. (\ref{Grad})] (solid lines) and Chapman-Enskog expansion
   [Eq. (\ref{CE})] (dashed lines) with $\tau_{\rm sw}=0.4$ fm/$c$.}
 \label{vnpt_t48CE_b23_LHC}
\end{figure}

Figure \ref{vnpt_t48CE_b23_LHC}(a) shows a comparison between
$v_n(p_T)$ obtained using two different switchover times, $\tau_{\rm
  sw}=0.4$ fm/$c$ (solid lines) and $\tau_{\rm sw}=0.8$ fm/$c$ (dashed
lines), for $20-30$\% central Pb+Pb collisions at $\sqrt{s_{NN}}=2.76$ TeV
at a fixed $\eta/s=0.12$. We find that within a window of $0.3 <
\tau_{\rm sw} \leq 0.8$ fm/$c$, a larger switchover time slightly
reduces $v_n(p_T)$, especially for lower harmonics. However, the
difference is negligible, which suggests that the flow buildup at
very early times is rather insensitive to the transport or
hydrodynamic modeling. This also indicates that the parton dynamics in
AMPT quickly brings the system to local equilibrium. The lack of
sensitivity to the switchover time was also noted in
Refs. \cite{Gale:2012rq,vanderSchee:2013pia}.

We now explore the uncertainty in the flow owing to a different choice
of the viscous correction in the nonequilibrium distribution function.
An alternate form to Grad's approximation, Eq. (\ref{Grad}), is based
on the Chapman-Enskog-like approach \cite{Bhalerao:2013pza},
\begin{equation}\label{CE}
 \delta f = \frac{5 f_0 \tilde f_0}{8 P T(u \!\cdot\! p)}\, p^\alpha p^\beta \pi_{\alpha\beta},
\end{equation}
which is obtained by iteratively solving the Boltzmann equation in the
relaxation-time approximation. It may be noted that, though the two
nonequilibrium distributions lead to distinct viscous corrections, the
dissipative evolution equations derived from them have identical forms
and coefficients in the extreme relativistic limit.
In Fig. \ref{vnpt_t48CE_b23_LHC}(b), we compare
$v_n(p_T)$ based on Grad's approximation (solid lines) and
Chapman-Enskog expansion (dashed lines) for the same initial
conditions. The two forms of $\delta f$ result in essentially similar
$v_n$ within the statistical errors. Note, however, that the
Chapman-Enskog results tend to be slightly below Grad's at small $p_T$
and slightly above Grad's at larger $p_T$. This is attributable to the fact that
the Chapman-Enskog method leads to essentially linear momentum 
dependence of $\delta f$ as opposed to the quadratic dependence in
case of Grad's approximation. This results in
larger (smaller) viscous corrections in the Chapman-Enskog approach at
smaller (larger) $p_T$ \cite{Bhalerao:2013pza}. In the following, we
use $\tau_{\rm sw}=0.4$ fm/$c$ for switchover time and Grad's
approximation for $\delta f$ in the freeze-out prescription.

Figure \ref{vnpt_b1234_LHC} shows our results for the rms $v_n(p_T)$,
in comparison with the ATLAS data \cite{ATLAS:2012at} obtained in the
event-plane method, at various centralities. We find overall good
agreement with the data for all harmonics ($n=2-6$) and at all
centralities. A single fixed value $\eta/s = 0.12$ of the shear
viscosity to entropy density ratio is able to achieve the required
suppression (relative to perfect hydrodynamics) of the flow harmonics
for all centralities. Apart from $\eta/s$, flow also depends on the
smearing parameter $\sigma$ appearing in Eq. (\ref{smear}). It
controls the granularity and hot spots in the initial state
\cite{Schenke:2011bn}. A smaller $\sigma$ results in a hardened $p_T$
spectrum owing to larger pressure gradients
\cite{Holopainen:2010gz}. Elliptic flow $v_2$ is less sensitive to
variations in $\sigma$, compared to the triangular flow $v_3$, which
is entirely driven by fluctuations. A smaller value of $\sigma$ tends
to raise the value of $v_3$, which can be compensated by choosing a
higher value of $\eta/s$. However, this would lead to underprediction
of $v_2$. This suggests that the initial-state fluctuations alone are
unlikely to reproduce all the observed $v_n$, and the interplay
between fluctuations and viscosity is essential. As several physical
sources of fluctuations in both the nucleons and partons contribute to
the AMPT initial conditions (as already described), a reasonably
diffused initial state with $\sigma = 0.8$ fm is required to explain
the $v_n ~(n=2-6)$ ATLAS data.

\begin{figure}[t]
 \begin{center}
  \scalebox{0.45}{\includegraphics{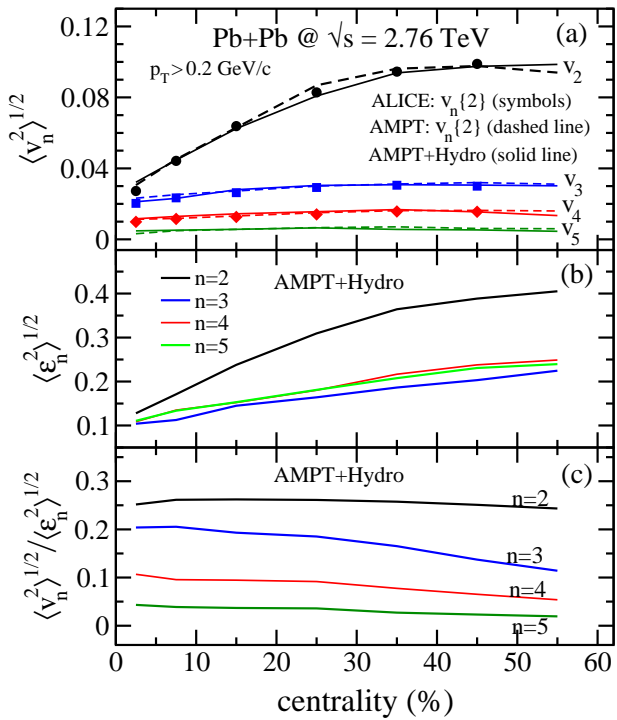}}
 \end{center}
 \vspace{-0.4cm}
 \caption{(Color online) (a) Centrality dependence of the rms values of the
   anisotropic flow coefficients $\langle v_n^2\rangle^{1/2}$ of
   charged hadrons in Pb+Pb collisions at $\sqrt{s_{NN}}=2.76$ TeV
   calculated in the AMPT+Hydro (solid lines) with $\eta/s =0.12$ as
   compared to $v_n\{2\}$ obtained from two-particle correlations within
   the complete AMPT calculations (dashed lines) and the ALICE data
   \cite{ALICE:2011ab} (symbols). (b) Centrality dependence of the rms
   values of the eccentricities $\langle \varepsilon_n^2\rangle^{1/2}$
   at the hydrodynamic switchover time. (c) Centrality dependence of
   the ratio $\langle v_n^2\rangle^{1/2}/\langle
   \varepsilon_n^2\rangle^{1/2}$ in the AMPT+Hydro model.}
 \label{vnen_cent_LHC}
\end{figure}

\begin{figure}[t]
 \begin{center}
 \scalebox{0.32}{\includegraphics{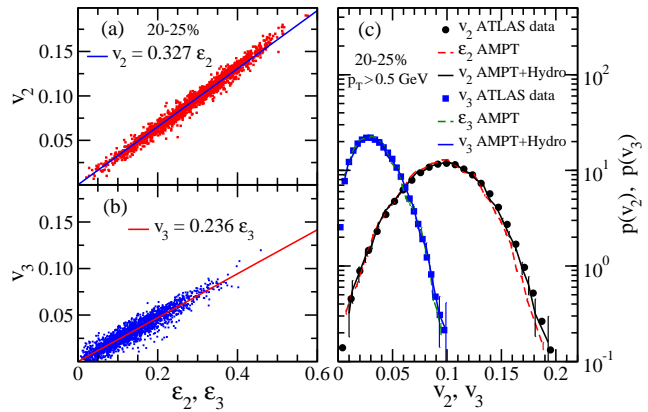}}
 \end{center}
 \vspace{-0.4cm}
 \caption{(Color online) (a),(b) Correlations between harmonic flow coefficients and
   eccentricities at $\tau=\tau_{\rm sw}$ for $n=2-3$. The solid
   lines represent the fits. (c) Event-by-event distributions $p(v_n)$
   calculated in AMPT+Hydro with $\eta/s =0.12$ (solid lines) and the
   eccentricity distributions, $p(\varepsilon_n)$ scaled to the
   respective $\mean{v_n}$ from AMPT initial conditions (dashed
   lines). Errors on the calculated distributions are shown.
   Symbols represent the ATLAS data
   \cite{Aad:2013xma}. Results are for charged hadrons at $p_T > 0.5$
   GeV/$c$ in $20-25\%$ central Pb+Pb collisions at $\sqrt{s_{NN}}=2.76$
   TeV. The analysis was performed with a sample of 3000 events.}
 \label{Pvn_LHC}
\end{figure}

In Fig. \ref{vnen_cent_LHC}(a), $p_T$-integrated rms values of $v_n$
in AMPT plus VISH2+1 calculations are compared with ALICE data
\cite{ALICE:2011ab} for $v_n\{2\}$ obtained from two-particle
correlations, as a function of centrality. 
Our calculation of the $p_T$-integrated flow takes into account the
Jacobian appearing in the transformation between rapidity and
pseudorapidity \cite{Kolb:2001yi}. It also takes into account
pseudorapidity and $p_T$ cuts appropriate for the ALICE data.
With the same constant $\eta/s =
0.12$ at all centralities, the model shows a remarkable agreement with
the data for all harmonics. As expected, $v_2$ exhibits a strong
centrality dependence as it is driven mostly by the initial spatial
anisotropy, whereas the odd harmonics are entirely induced by
fluctuations.

Figure \ref{vnen_cent_LHC}(a) also shows the $v_n\{2\}$ obtained
solely in the AMPT model without the inclusion of VISH2+1 hydrodynamic
evolution. This is slightly larger than that obtained in the
AMPT+Hydro calculation. This can be understood by noting that the
hybrid calculation employs a softer EoS with a crossover transition
and a larger viscous damping, relative to the pure AMPT calculation
\cite{Zhang:2008zzk,Pal:2009zz}.

Figure \ref{vnen_cent_LHC}(b) shows the rms value of the eccentricity
at $\tau_{\rm sw}=0.4$ fm/$c$, as a function of centrality; see
Eq. (\ref{eccn}). In the limit of ultracentral collisions, the
eccentricities arise solely from fluctuations and are found to be
nearly identical. As the impact parameter increases, they all
increase: The ellipticity becomes more pronounced for midcentral
collisions, and for the higher harmonics the fluctuations become more
important for smaller overlapping geometry. However, as expected,
$\varepsilon_2 > \varepsilon_3 \approx \varepsilon_4 \approx
\varepsilon_5$ as the shape of the overlap zone is predominantly
elliptic.

We now study the efficiency of conversion of the initial spatial
anisotropy into the final momentum anisotropy as a result of
hydrodynamic evolution. To that end, we present in
Fig. \ref{vnen_cent_LHC}(c) the ratio $k_n$ of rms values of $v_n$
and $\varepsilon_n$. Given the nonlinear nature of the hydrodynamic
equations of motion, a linear relation between $v_n$ and
$\varepsilon_n$ is {\it a priori} not obvious. However, a strong linear
correlation is seen in Fig. \ref{vnen_cent_LHC}(c), especially for
$n=2$, as in several other initial-state models. We find that at a
given centrality, the ratio $k_n$ decreases as $n$ increases,
indicating reduced conversion efficiency for higher harmonics. As
stated above, the shape of the overlap zone is predominantly elliptic,
which makes the conversion of $\varepsilon_2$ into $v_2$ relatively
more efficient than that for higher harmonics. For $n > 2$, the shapes
are governed by small-scale structures which makes the conversion
relatively less efficient.  For $n > 2$ and peripheral collisions, the
short lifetime of the system makes the conversion even weaker.

Fluctuations in the initial positions of the nucleons and the formed
partons in an event lead to an event-by-event distribution in the
initial $\varepsilon_n$ as well as in the final $v_n$. Figures
\ref{Pvn_LHC}(a) and \ref{Pvn_LHC}(b) show the scatter plots for $v_2$ versus
$\varepsilon_2$ and $v_3$ versus $\varepsilon_3$ for $20-25\%$
centrality Pb+Pb collisions in the AMPT+Hydro model. Note that
$\varepsilon_n$ are calculated at $\tau=\tau_{\rm sw}$ using
Eq. (\ref{eccn}). The scatter plots can be described to a good
approximation by a linear relation $\langle v_n \rangle = k_n
\varepsilon_n$, where $k_2=0.327$ and $k_3=0.236$. Figure
\ref{Pvn_LHC}(c) shows the probability density distribution $p(v_n)$
for charged particles with $p_T > 0.5$ GeV/$c$ at the same centrality in
this model and compares it with the ATLAS data
\cite{Aad:2013xma}. The shapes and magnitudes of the $v_2$ and $v_3$
distributions are remarkably well reproduced by the model. We also
show in Fig. \ref{Pvn_LHC}(c) the scaled probability density
distribution $p(\varepsilon_n)/k_n$ for the AMPT initial collision
geometry. These too are in excellent agreement with the corresponding
$p(v_n)$ distributions, except at large $v_2$, where the scaled
distribution for $n=2$ drops faster. This is attributable to slightly
enhanced conversion coefficient $k_2 = \mean{v_2}/\varepsilon_2$ at large
$\varepsilon_2$, as can be seen in Fig. \ref{Pvn_LHC}(a). This deviation
from the linear behavior has been also observed within recent
hydrodynamic model simulations \cite{Niemi:2014ita}.

\begin{figure}[t]
 \begin{center}
 \scalebox{0.4}{\includegraphics{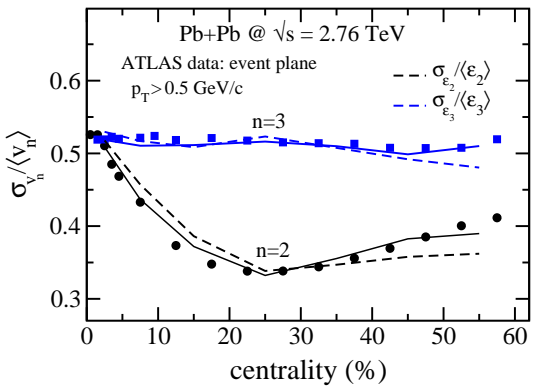}}
 \end{center}
 \vspace{-0.4cm}
 \caption{(Color online) Centrality dependence of the ratio $\sigma_{v_n}/\langle
   v_n\rangle$, for charged hadrons with momenta $p_T > 0.5$ GeV/$c$ in
   the AMPT+Hydro model (solid lines) as compared with the ATLAS data
   \cite{Aad:2013xma} (symbols). Also shown is the ratio
   $\sigma_{\varepsilon_n}/\langle \varepsilon_n\rangle$ obtained in
   the AMPT model (dashed lines).}
 \label{sigbyvn}
\end{figure}

These distributions can be used to calculate the respective mean and
variance. In Fig. \ref{sigbyvn} we present the centrality dependence
of $\sigma_{v_n}/\langle v_n \rangle$ and
$\sigma_{\varepsilon_n}/\langle \varepsilon_n \rangle$ for $n=2,3$ in
the AMPT+Hydro calculations. These results are compared with the ATLAS
data for $\sigma_{v_n}/\langle v_n \rangle$ \cite{Aad:2013xma}. In the
fluctuations-only scenario, these ratios are expected to be about 0.52
\cite{Aad:2013xma}, which is corroborated by the model calculations
for $n=2$ in the ultracentral limit and for $n=3$ at all
centralities.
With increasing impact parameter, the smaller number of participants
causes the event-by-event fluctuations and hence the variance
$\sigma_{\varepsilon_n}$ to increase gradually. Now, for $n=2$, the
$\mean{\varepsilon_2}$ increases rapidly up to $\sim 25-30\%$
centrality, and thereafter the increase is somewhat slower; see
Fig. \ref {vnen_cent_LHC}(b). Consequently, the ratio
$\sigma_{\varepsilon_2}/\langle \varepsilon_2 \rangle$ shows a minimum
at $\sim 25-30\%$ centrality and a slower growth rate at higher
centralities. Similar behavior is seen for the ratio
$\sigma_{v_2}/\mean{v_2}$, which seems to support the linear
hydrodynamic response, at least up to midcentral collisions.
For peripheral collisions, the failure of the linear hydrodynamic
response ($v_n \propto \varepsilon_n$), as discussed above, leads to
the disagreement between the two ratios.

\begin{figure}[t]
 \begin{center}
  \scalebox{0.45}{\includegraphics{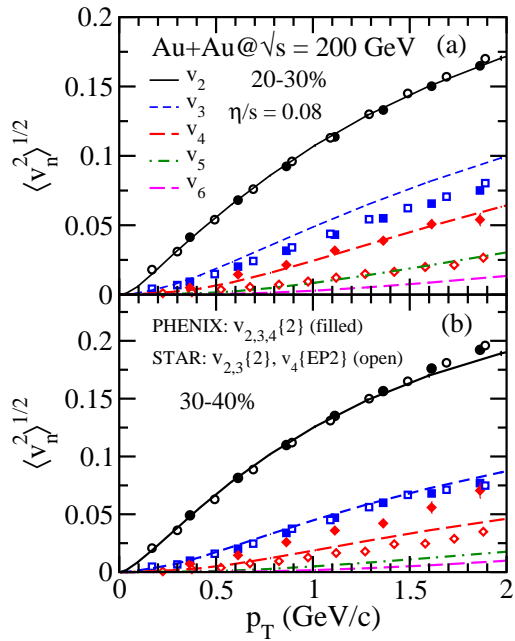}}
 \end{center}
 \vspace{-0.4cm}
 \caption{(Color online) Transverse-momentum dependence of $\langle
   v_n^2(p_T)\rangle^{1/2}$ for charged hadrons in $20-30\%$ and
   $30-40\%$ central Au+Au collisions at $\sqrt{s_{NN}}=200$ GeV in the
   AMPT+Hydro calculations with $\eta/s=0.08$ (lines) as compared with
   the experimental data from PHENIX \cite{Adare:2011tg} (solid
   symbols) and STAR \cite{Adams:2004bi,Adamczyk:2013waa} (open
   symbols).}
 \label{vnpt_b14_RHIC}
\end{figure}

Finally, for the sake of completeness, we show in
Fig. \ref{vnpt_b14_RHIC} the rms values of $v_n(p_T)$ in the full
AMPT+Hydro calculation in comparison with the PHENIX and STAR data for
Au+Au collisions at the top RHIC energy.  With the shear viscosity to
entropy density ratio of $\eta/s = 0.08$, which corresponds to $33\%$
reduction from the LHC value 0.12 estimated above, the model describes
the RHIC data well within the systematic uncertainties. A smaller
value of $\eta/s$ at RHIC compared to that at LHC was also found in
previous studies \cite{Gale:2012rq}
that employed various fluctuating initial conditions.

\section{Conclusions}

In summary, within a coupled AMPT transport plus (2+1)D viscous
hydrodynamics approach, we have studied various aspects of the
anisotropic flow coefficients in heavy-ion collisions at RHIC and
LHC. Fluctuations in both the number and positions of the participant
nucleons and the formed partons and the subsequent parton dynamics, in
the event-by-event AMPT model, provide realistic
initial conditions for the later viscous evolution of the QGP
and hadronic matter until it decouples. We find that this model
provides a good agreement with the transverse-momentum-dependent and
integrated flow coefficients $v_n ~(n=2-6)$ at RHIC with a minimal
$\eta/s=1/4\pi$ and at LHC with a larger value $\eta/s = 1.5/4 \pi$ at
various collision centralities.

Our other results are as follows.
We have isolated the contributions of parton cascade to the generation
of initial eccentricities and the collective flow, and they are found
to be significant.
Insensitivity of the results to the precise value of the switchover
time (within a window) suggests that the partonic scatterings in the AMPT
model drive the system to near equilibrium in less than $\sim 1$ fm/$c$.
For ultracentral collisions, the model is able to reproduce
qualitatively the unexpected hierarchy of $v_n(p_T)$ observed by CMS.
The model agrees with the observed probability distributions $p(v_2)$
and $p(v_3)$ at midcentral collisions.
There is a hint of a nonlinear hydrodynamic response even for the
harmonics $n=2,3$ at large eccentricities.

\section{Acknowledgments}

We thank U. Heinz, Z. Qiu, and C. Shen for useful correspondence on
the VISH2+1 code. A.J. was supported by Frankfurt Institute for
Advanced Studies (FIAS), Germany.

\end{document}